\documentclass[traditabstract]{aa}
\usepackage{txfonts}
\usepackage{graphicx}
\usepackage{subfig}
\usepackage{natbib}
\usepackage{subfig}
\bibpunct{(}{)}{;}{a}{}{,}

\usepackage[colorlinks=true, citecolor=blue, linkcolor=blue, breaklinks=true]{hyperref}
\newcommand{\hi } {{\rm H}\,{\small\rm I} \,}
\newcommand{\hii } {{\rm H}\,{\small\rm II} \,}
\newcommand{\hiV} {{\rm H}\,{\small\rm I}}
\newcommand{\kms} {km~s$^{-1}$ \,}
\begin{document}

\title{Dynamics of starbursting dwarf galaxies. II. UGC~4483}
\author{Federico Lelli\inst{1}
\and Marc Verheijen\inst{1}
\and Filippo Fraternali\inst{1,}\inst{2}
\and Renzo Sancisi\inst{1,}\inst{3}}

\institute{Kapteyn Astronomical Institute, University of Groningen, Postbus 800, 9700 AV, Groningen, The Netherlands \\
\email{lelli@astro.rug.nl}
\and Department of Astronomy, University of Bologna, via Ranzani 1, 40127, Bologna, Italy
\and INAF - Astronomical Observatory of Bologna, via Ranzani 1, 40127, Bologna, Italy}

\date{}

\abstract{
UGC~4483 is a nearby Blue Compact Dwarf (BCD) galaxy. HST observations have resolved the
galaxy into single stars and this has led to the derivation of its star formation history
and to a direct estimate of its stellar mass. We have analysed archival VLA observations
of the 21-cm line and found that UGC~4483 has a steeply-rising rotation curve which
flattens in the outer parts at a velocity of $\sim$20~km~s$^{-1}$. Radial motions
of $\sim$5 km~s$^{-1}$ may also be present. As far as we know, UGC~4483 is the lowest-mass
galaxy with a differentially rotating \hi disk. The steep rise of the rotation curve
indicates that there is a strong central concentration of mass. We have built mass models
using the HST information on the stellar mass to break the disk-halo degeneracy: old stars
contribute $\sim$50$\%$ of the observed rotation velocity at 2.2 disk scale-lengths.
Baryons (gas and stars) constitute an important fraction of the total dynamical mass. These are
striking differences with respect to typical dwarf irregular galaxies (dIrrs), which usually have
slowly-rising rotation curves and are thought to be entirely dominated by dark matter. BCDs
appear to be different from non-starbursting dIrrs in terms of their \hi and stellar distributions
and their internal dynamics. To their high central surface brightnesses and high central \hi
densities correspond strong central rotation-velocity gradients. This implies that the
starburst is closely related with the gravitational potential and the concentration of gas.
We discuss the implications of our results on the properties of the progenitors/descendants of BCDs.
}

\titlerunning{Dynamics of starbursting dwarf galaxies. II. UGC~4483.}
\authorrunning{Lelli et al.}

\maketitle

\section{Introduction}

The mechanisms that trigger strong bursts of star formation in galaxies are poorly understood.
In the Local Universe, starburst activity is mostly observed in low-mass galaxies, which
are usually classified as blue compact dwarfs (BCDs) (e.g. \citealt{GilDePaz2003}),
amorphous dwarfs (e.g. \citealt{Gallagher1987}), or \hii galaxies (e.g. \citealt{Taylor1995}).
Hereafter, we will refer to any starbursting dwarf galaxy as a BCD. Several studies
(e.g. \citealt{GilDePaz2005}, \citealt{Tosi2009} and references therein) have shown that BCDs
are \textit{not} young galaxies undergoing their first burst of star formation (as suggested
by \citealt{Searle1972}), as they also contain old stellar populations with ages $>$2-3~Gyr.
In particular, HST has made it possible to resolve nearby BCDs into single stars and to derive
colour-magnitude diagrams deep enough to provide the following information: i) accurate
distances of the galaxies, ii) a direct estimate of their total stellar mass, and iii)
their star formation history (SFH) (e.g. \citealt{Tosi2009}).
These SFHs show that the starburst is a short-lived phenomenon, typically sustained for
a few 10$^{8}$~yr (\citealt{McQuinn2010}). Thus, BCDs are \textit{transition-type dwarfs}
but the nature of their progenitors and descendants remains unclear. In particular, it is not
known whether there are evolutionary connections with dwarf irregulars (dIrrs), spheroidals
(dSphs), and/or ellipticals (dEs) (e.g. \citealt{Papaderos1996, vanZee2001}).

There are striking differences between BCDs and other types of dwarf galaxies: i) the old stellar
component of BCDs generally has a smaller scale-length and higher central surface brightness than
dIrrs and dEs/dSphs (e.g. \citealt{Papaderos1996, GilDePaz2005}); ii) BCDs have strong concentrations
of \hi within the starburst region, where the column densities are typically 2-3 times higher than in
dIrrs (e.g. \citealt{vanZee1998b, vanZee2001}); iii) BCDs have steep central velocity gradients that
are not observed in dIrrs (e.g. \citealt{vanZee1998b, vanZee2001}). The steep velocity gradients may
signify a steeply-rising rotation curve \citep{vanZee2001, Lelli2012}, high velocity dispersion,
or non-circular motions (e.g. \citealt{Elson2011b}). Detailed studies of the gas kinematics are
needed to determine the inner shape of the rotation curve. Recently, \citet{Lelli2012} studied
the BCD prototype I~Zw~18 and found that it has a flat rotation curve with
a steep rise in the inner parts, indicating that there is a high central concentration of mass.
Such a mass concentration is not observed in typical dIrrs. This points to a close connection
between the starburst and the gravitational potential. It is also clear that a BCD like I~Zw~18 cannot
evolve into a typical dIrrs at the end of the starburst, unless the central concentration of mass is
removed. It is important, therefore, to investigate whether all BCDs have steeply-rising rotation curves,
and to determine the relative contributions of gas, stars and dark matter to the gravitational potential.

\begin{figure*}[tbp]
\centering
\includegraphics[width=17.4cm]{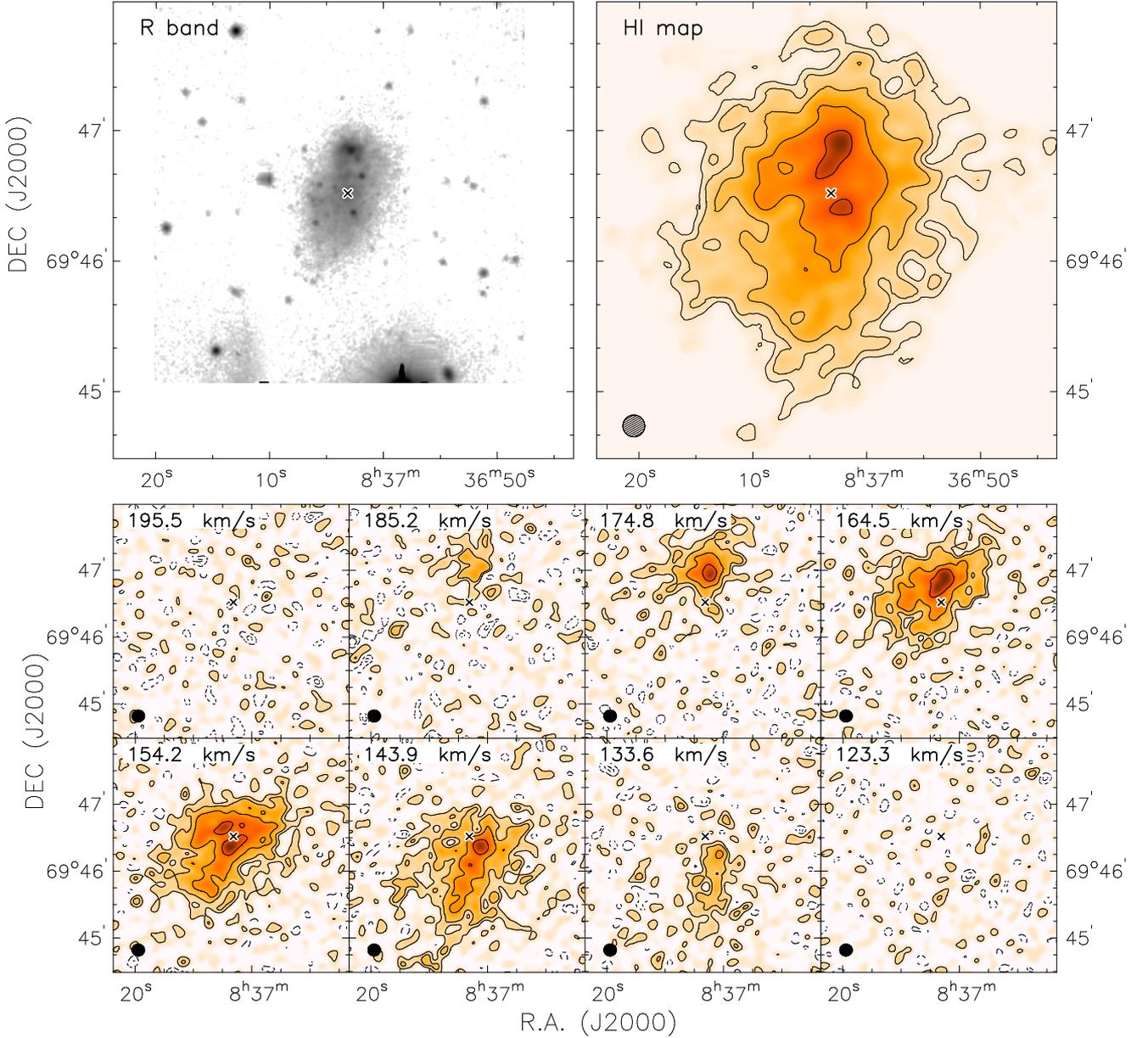}
\caption{\textit{Top-Left:} R-band image (from \citealt{GilDePaz2003}).
\textit{Top-Right:} total \hi map at 10$''$ resolution. The contour values are 1.7, 3.4,
6.8, 13.6, 27.2 $\times$ 10$^{20}$ atoms~cm$^{-2}$. \textit{Bottom:} channel maps at
10$''$ resolution. Contours are at 1.5, 3, 6, 12, 24~$\times$~$\sigma$, where $\sigma$=0.66
mJy/beam ($\sim$2$\times$10$^{19}$ atoms~cm$^{-2}$). In each panel, the cross and the
circle show the galaxy centre and the beam size, respectively.}
\label{fig:mosaic}
\end{figure*}

\begin{table}[thp]
\caption{Properties of UGC~4483.}
\label{tab:prop}
\centering
\begin{tabular}{l c c}
\hline
\hline
$\alpha$ (J2000)                      & $08^{\rm{h}} 37^{\rm{m}} 3.^{\rm{s}}1 \pm 0.^{\rm{s}}5$ \\
$\delta$ (J2000)                      & $69^{\circ} 46' 31'' \pm 2''$                   \\
Distance (Mpc)                        & $3.2 \pm 0.2$                                   \\
$V_{\rm{sys}}$ (km s$^{-1}$)          & $158 \pm 2$                                  \\
Position Angle ($^{\circ}$)           & $0   \pm 5$                                     \\
Inclination Angle ($^{\circ}$)        & $58  \pm 3$                                     \\
$V_{\rm{rot}}$ (km s$^{-1}$)          & $19  \pm 2$                                   \\
$M_{\rm{dyn}}$ (10$^{7}$ $M_{\odot}$) & $16 \pm 3$                                       \\
$M_{*}$ (10$^{7}$ $M_{\odot}$)        & $1.0 \pm 0.3$                                     \\
$M_{\hi}$ (10$^{7}$ $M_{\odot}$)      & $2.5 \pm 0.3$                                   \\
$L_{\rm{B}}$ (10$^{7}$ $L_{\odot}$)   & 1.4                                            \\
$L_{\rm{R}}$ (10$^{7}$ $L_{\odot}$)   & 0.9                                             \\
\hline
\end{tabular}
\tablefoot{Luminosities were calculated using the apparent magnitudes
from \citet{GilDePaz2003}, the distance from \citet{Dolphin2001} and the solar
absolute magnitudes from \citet{BinneyMerrifield1998}. The stellar mass
was calculated integrating the SFH from \citet{McQuinn2010} and assuming a gas
recycling efficiency of 30$\%$. The dynamical mass was calculated taking into
account the pressure-support.}
\end{table}
We present a detailed study of the gas kinematics of UGC~4483, a starbursting dwarf galaxy
located in the M81 group and resolved into individual stars by HST
\citep{Dolphin2001, Izotov2002}. UGC~4483 is extremely metal-poor (12+log(O/H)$\simeq$7.5,
see \citealt{Skillman1994} and \citealt{vanZee2006}) and may be classified as a ``cometary''
BCD, as its high-surface-brightness starburst region is located at the edge of an elongated
low-surface-brightness stellar body (see Fig.~\ref{fig:mosaic}, top-left). Previous \hi studies
(\citealt{Lo1993, vanZee1998b}) showed that the galaxy has an extended \hi disk, with a strong
\hi concentration near the starburst region and a steep central velocity gradient, but the \hi
kinematics was not studied in detail. We analysed archival VLA data and were able to derive
a rotation curve which we used to investigate the distributions of luminous and dark matter
in this galaxy.

\section{Data reduction \& analysis \label{sec:reduction}}

We analysed \hi data taken from the VLA archive. The observations have been carried
out with the B and C arrays, and are described in \citet{vanZee1998b}. The correlator was
used in 2AD mode, with a total bandwidth of 1.56 MHz ($\sim$330 km s$^{-1}$). An on-line
Hanning taper was applied to the data, producing 127 spectral line channels with a width
of 12.3 kHz ($\sim$2.6 km s$^{-1}$).

We interactively flagged, calibrated, and combined the raw UV data using the AIPS
package and following standard VLA procedures. The UV data were mapped using a robust
weighting technique \citep{Briggs1995} and a Gaussian baseline taper to attenuate the
longest baselines. After various trials, we chose a robust parameter of -1 and a taper
FWHM of 40 k$\lambda$; these parameters minimize sidelobes and wings in the beam profile
and lead to a datacube with an angular resolution of $5''.7\times4''.5$.

After the Fourier transform, the data analysis was continued using the Groningen Imaging
Processing SYstem (GIPSY) \citep{vanderHulst1992}. A continuum map was constructed by
averaging line-free channels, and subtracted from the datacube. The channel maps were
cleaned \citep{Hogbom1974} down to 0.3$\sigma$, using a mask to define the search
areas for the clean-components that were then restored with a Gaussian beam of the
same FWHM as the antenna pattern. The mask was constructed by smoothing the datacube
both in velocity (to 10.4 km~s$^{-1}$) and spatially (to 20$''$) and clipping at
$\sim$3$\sigma_{s}$ (where $\sigma_{s}$ is the rms noise in the smoothed cube).
In order to improve the S/N, the cleaned datacube was smoothed in velocity to a resolution
of 5.2 km~s$^{-1}$ and spatially to 10$''$, providing a 3$\sigma$ column density
sensitivity of 6$\times$10$^{19}$ atoms~cm$^{-2}$ per 2.6~km~s$^{-1}$-wide channel.

A total \hi map was constructed by summing the signal inside the clean-mask;
a pseudo-3$\sigma$ contour was calculated following \citet{Verheijen2001}.
A velocity field was derived by fitting a Gaussian function to the \hi line profiles.
Fitted Gaussians with a peak intensity less than 3$\sigma$ and a FWHM smaller
than 5.2 \kms were discarded. The \hi line profiles are quite broad and asymmetric,
thus the velocity field derived from the Gaussian fitting provides only an overall
description of the galaxy kinematics. Our kinematical analysis is based on
three-dimensional (3D) models of the observations (Sect. \ref{sec:kin}) and
not merely on the two-dimensional (2D) velocity field.

\section{Results \label{sec:results}}

\subsection{\hi distribution and kinematics \label{sec:HIkin}}

Figure \ref{fig:mosaic} shows the total \hi map of UGC~4483 at a resolution
of 10$''$ (top-right) and a R-band image at the same scale (top-left).
The \hi distribution is lopsided and closely resembles the optical morphology.
There is a strong \hi concentration near the starburst region to the North.
At 5.$''$7$\times$4$''$.5 resolution ($\sim$80~pc), the peak column densities
are $\sim$5$\times$10$^{21}$ atoms~cm$^{-2}$ ($\sim$40 $M_{\odot}$~pc$^{-2}$)
(same as found by \citealt{vanZee1998b}).

The \hi kinematics of UGC~4483 is illustrated in Fig.~\ref{fig:mosaic} (bottom) and
Fig.~\ref{fig:velo} (top). The \hi emission shows a velocity gradient along a position
angle P.A.$\sim$0$^{\circ}$, which roughly corresponds to the optical major axis of
the galaxy, suggesting that there is a rotating \hi disk. The velocity field indicates
large-scale differential rotation. It also shows, however, large-scale asymmetries. 
In the next section, we derive the rotation curve of
UGC~4483 and discuss possible non-circular motions.

\subsection{Rotation curve\label{sec:kin} and non-circular motions}

Rotation curves of disk galaxies are usually derived by fitting a tilted-ring model to a velocity
field (e.g. \citealt{Begeman1987}). In the case of UGC~4483, there are severe limitations because
of the asymmetries in the velocity field. Van Zee et al. (1998) modeled the velocity field of
UGC~4483 and obtained a rough estimate of the dynamical mass. We derived a rotation curve using the
following approach. As a first step, we fit a tilted-ring model to the velocity field to
obtain an initial estimate of the rotation curve. Then, this rotation curve was used as input to
build a 3D kinematic model and subsequently corrected by trial and error to produce a model-cube
that matches the observations.

\begin{figure*}[tbp]
\centering
\includegraphics[width=17cm]{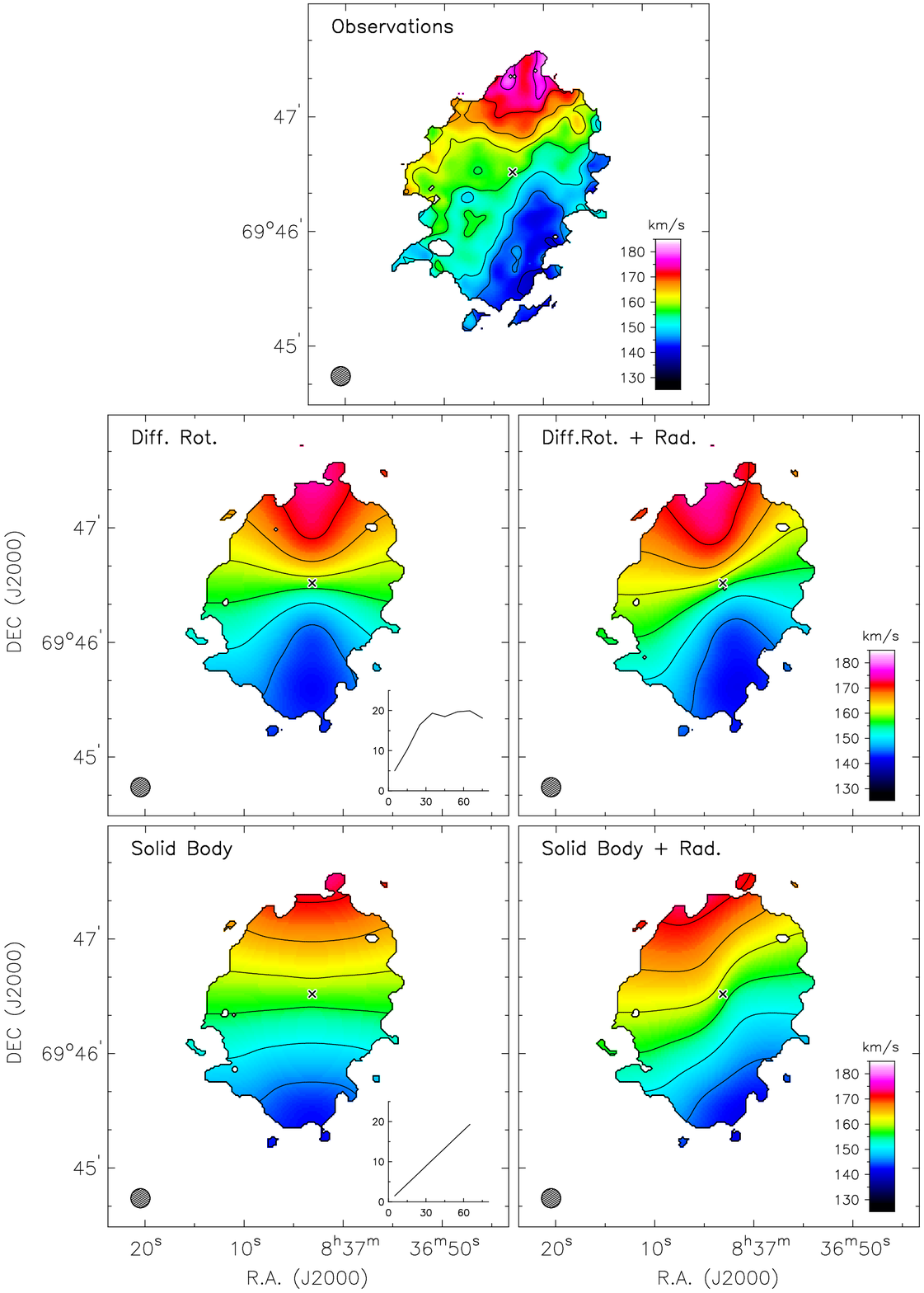}
\caption{Velocity fields derived from different 3D kinematic models and the observations.
The rotation curve used to build the models is shown in the inset on the left (x-axis: radius
in arcsec, y-axis: rotation velocity in km~s$^{-1}$). Contours range from 140 to 181.6 km~s$^{-1}$,
with steps of 5.2~km~s$^{-1}$. The circles show the beam size (10$''$). See Sect.~\ref{sec:HIkin} for details.}
\label{fig:velo}
\end{figure*}
When fitting a tilted-ring model to the velocity field, we used a ring width of $10''$ (1 beam), thus the points
of the rotation curve are nearly independent. The points of the velocity field were weighted by $cos^{2}(\theta)$,
where $\theta$ is the azimuthal angle in the plane of the galaxy. We kept the centre $(x_{0}, y_{0})$ and
the inclination $i$ fixed, using the values derived from the optical image by fitting ellipses to the
outermost isophotes (see table \ref{tab:prop}). An inclination of 58$^{\circ}$ is formally
a lower limit, because the stellar component may be thick; if the disk is assumed to be 5$^{\circ}$
more edge-on, the rotation velocities would decrease by only $\sim$5\%. We also fixed the P.A.
assuming the value of 0$^{\circ}$ suggested by the \hi morphology (Fig.~\ref{fig:mosaic}, top),
which is consistent with the optical P.A. ($\simeq-10^{\circ}$) within the uncertainties.
As a first step, we determined the systemic velocity $V_{\rm{sys}}$, by taking the mean value over all the rings.
Then, we fixed $V_{\rm{sys}}$ and determined the rotation velocity $V_{\rm{rot}}$ and the radial velocity
$V_{\rm{rad}}$ for each ring. The rotation curve has an amplitude of $\sim$18-20 km~s$^{-1}$
(Fig. \ref{fig:asymDrift}), while the radial velocities are $\sim$4-5~km~s$^{-1}$.

\begin{figure*}[tbp]
\centering
\includegraphics[width=18cm]{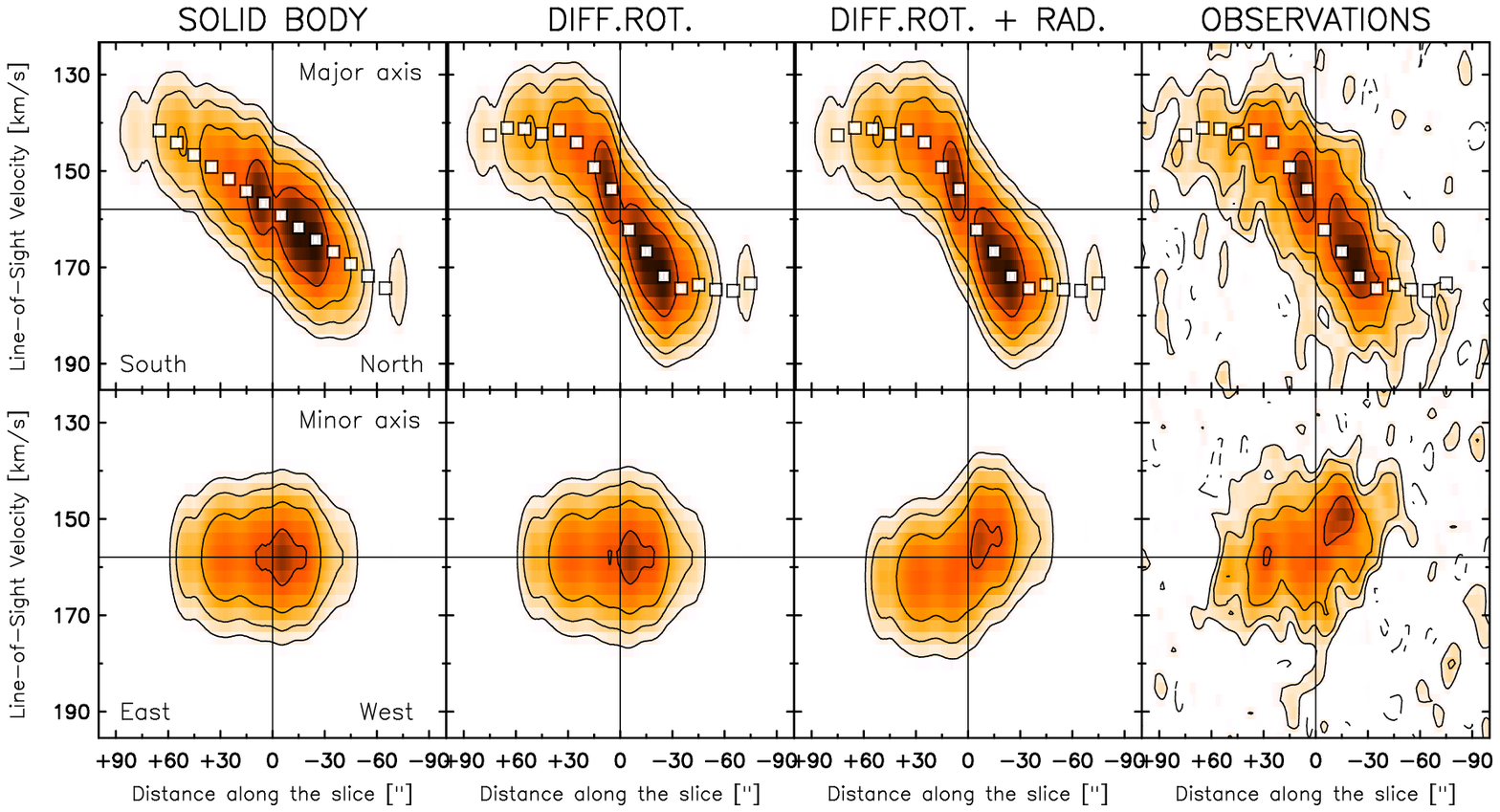}
\caption{Position-velocity diagrams derived for different 3D kinematic models and from the observations.
The slices are taken along the major and minor axes. Contours are at 1.5, 3, 6, 12 $\times$ $\sigma$, where
$\sigma$=0.66 mJy/beam. Squares show the rotation curve used to build the models. See Sect.~\ref{sec:HIkin}
for details.}
\label{fig:PVmod}
\end{figure*}
\begin{figure}[h!]
\centering
\includegraphics[width=8cm, angle=-90]{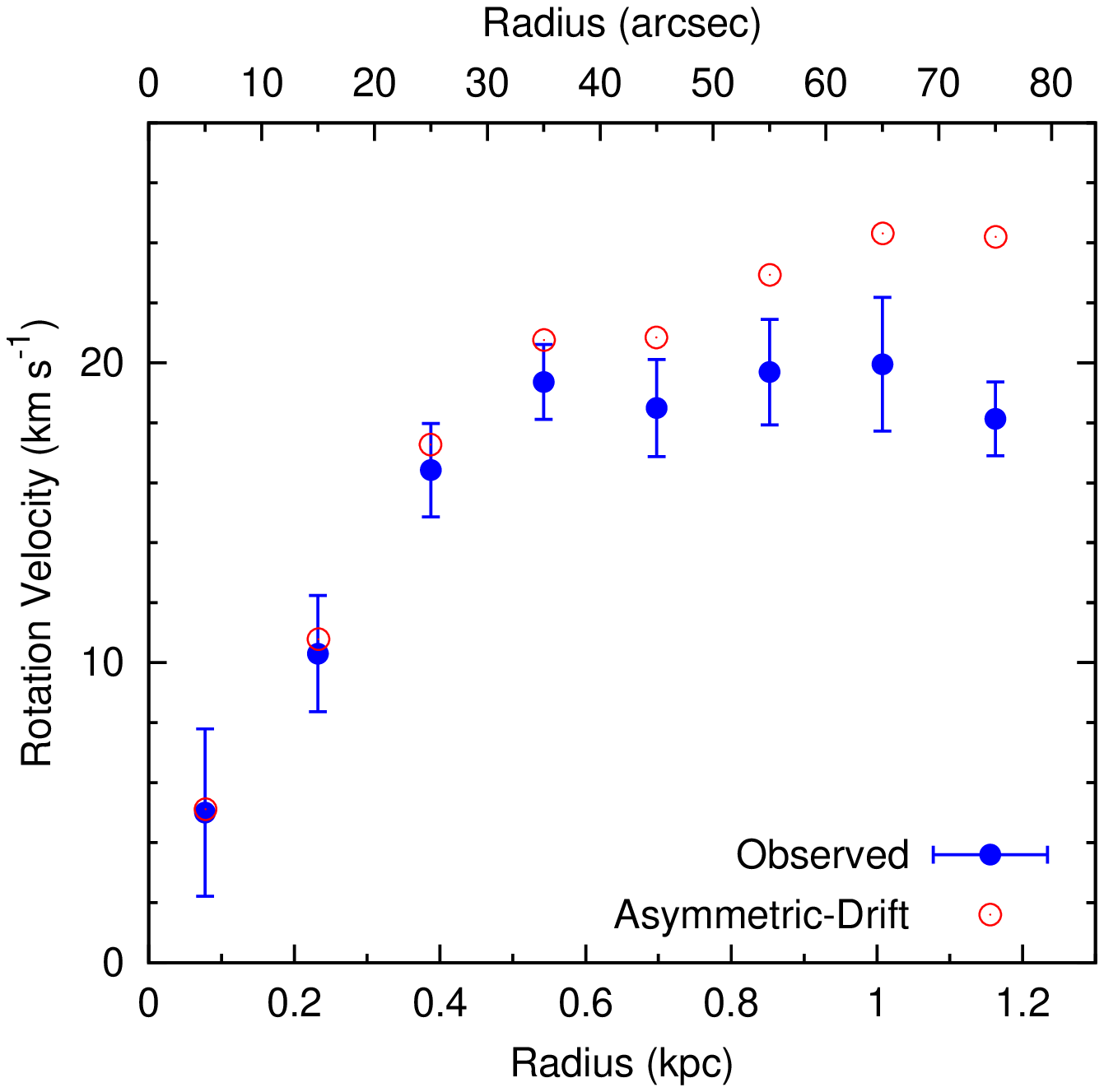}
\caption{Observed \hi rotation curve (blue-filled circles) and asymmetric-drift-corrected
rotation curve (red-open circles).}
\label{fig:asymDrift}
\end{figure}
As a final step in the derivation of the rotation curve, we built 3D kinematic models, similarly to
\citet{Swaters2009} and \citet{Lelli2010, Lelli2012}. The disk kinematics is assumed to be axisymmetric, while
the \hi distribution is clumpy, i.e. the surface density varies with position as in the observed \hi map.
The procedure is as follows: i) a disk with \textit{uniform} surface density and constant thickness is constructed
by fixing the velocity dispersion $\sigma_{\hi}$, the rotation velocity $V_{\rm{rot}}$ and the radial velocity
$V_{\rm{rad}}$ at every radius; ii) the disk is projected on the sky using the geometrical parameters
$(x_{0}, y_{0})$, V$_{\rm{sys}}$, P.A., and $i$ and a model-cube is created; iii) the model-cube
is convolved with the observational beam; and iv) the \hi line profiles are rescaled to reproduce
the observed \hi map, i.e. the flux density is recovered at every spatial pixel.
For the geometrical parameters, we used the values in table \ref{tab:prop}.
For the vertical distribution, we assumed an exponential-law $exp(-z/z_{0})$ with $z_{0}$=100~pc.
We also assumed that $\sigma_{\hi}$=8 km~s$^{-1}$ over the entire disk; a mean velocity dispersion
higher than 10 km~s$^{-1}$ can be ruled out by comparing the shape of position-velocity (PV)
diagrams obtained from the models and the observations. The values of $z_{0}$ and $\sigma_{\hi}$
are slightly degenerate, but do not significantly affect the final results. For the rotation curve,
we used the rotation velocities obtained by fitting the velocity field, then we corrected them
by trial and error to obtain a model-cube that matches the observations. We find a good match
by increasing the first four points by $\sim$3 km~s$^{-1}$. This rotation curve is shown in
Fig.~\ref{fig:asymDrift}. The comparison between models and observations has been done using
velocity fields (see Fig.~\ref{fig:velo}) and PV-diagrams (see Fig.~\ref{fig:PVmod}), 
Note that the model velocity fields were derived from the model-cubes and, therefore, 
include the effects of spatial and spectral resolution, velocity dispersion, and possible
non-circular motions. Because in our models we initially assumed a uniform \hi disk, some aspects
of the effects of beam-smearing are not fully included if the \hi distribution varies rapidly on
small scales. However, we used PV-diagrams at the full resolution (5.$''$7$\times$4$''$.5) and
compared them with those at 10$''$ to check that beam-smearing effects do not significantly
affect our rotation curve.

Figure~\ref{fig:asymDrift} (filled-circles) shows that UGC~4483 has a steeply-rising rotation curve
that flattens in the outer parts at a velocity of $\sim$18-20 km~s$^{-1}$. This is a striking difference with
respect to dIrrs of similar mass, as they usually have a slowly-rising rotation curve close to a solid body.
As far as we know, UGC~4483 is the lowest-mass galaxy with a differentially rotating \hi disk.
To further test the validity of this result, we built 3D kinematic models assuming a solid-body
rotation curve and compared them with the differentially-rotating disk-model and with the observations.
The results are shown in Fig.~\ref{fig:velo} and Fig.~\ref{fig:PVmod}. It is clear that a flat
rotation curve provides a better match of the observations than a solid-body one, as the iso-velocity
contours of the observed velocity field display a curvature that is typical of differential rotation.
The same conclusions are reached by comparing PV-diagrams obtained from the models and the observations
along the major axis (Fig.~\ref{fig:PVmod}, top). The observed PV-diagram shows a ``flattening'' on
the southern-approaching side of the disk, that is reproduced only by a differentially rotating
disk. On the northern-receding side, the ``flattening'' is less clear as the disk is
less extended, but a differentially rotating disk is still preferable than a solid body one.

\begin{figure}[h!]
\centering
\includegraphics[width=8.5cm]{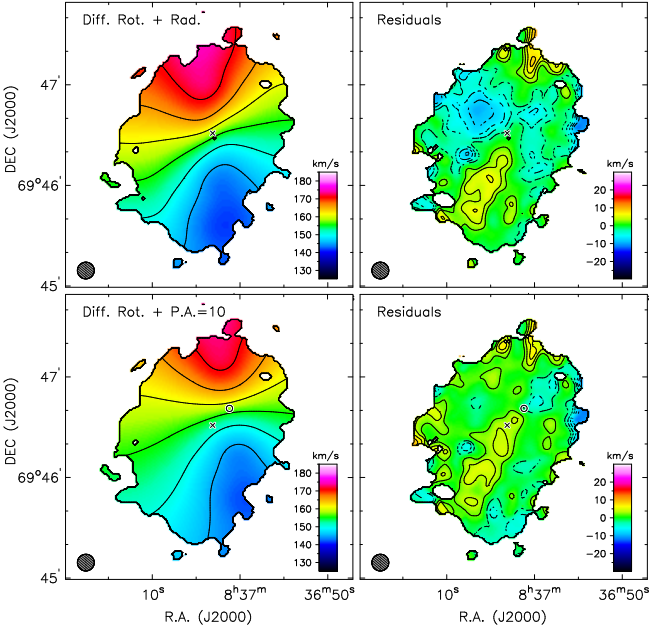}
\caption{\textit{Left}: Velocity fields derived from disk-models with different geometrical parameters:
optical centre (cross), P.A.$=$0$^{\circ}$, and radial motions of $\sim$5 km~s$^{-1}$ (top panel,
same as Fig.~\ref{fig:velo} middle-right); shifted dynamical centre (circle), P.A.$=$10$^{\circ}$,
and no radial motions (bottom panel). Contours are the same as in Fig.~\ref{fig:velo}.
\textit{Right:} differences between the observed and the model velocity fields. Contours
are at $\pm$2, $\pm$4, $\pm$6, $\pm$8 km~s$^{-1}$. The filled circle shows the beam.}
\label{fig:residuals}
\end{figure}
A simple rotating disk, however, cannot reproduce the asymmetries present in the observed
velocity field. In particular, the kinematic minor axis, defined by the contours
close to the systemic velocity, is not orthogonal to the major axis. This may be due either
to radial motions (e.g. \citealt{Fraternali2002}) or to an oval distortion of the gravitational
potential (e.g. \citealt{Bosma1978}). These asymmetries can be seen also in the PV-diagram
taken along the \textit{geometrical} minor axis, i.e. in the direction perpendicular to the
major axis (Fig~\ref{fig:PVmod}, bottom). Thus, we improved our models
by adding a constant radial component of $\sim$5 km~s$^{-1}$, which is also indicated
by the tilted-ring fit to the velocity field. A model with solid-body rotation
plus radial motions is still not acceptable (Fig.~\ref{fig:velo}, bottom-right), whereas
a model with differential rotation plus radial motions reproduces most of the features
in the data (Fig.~\ref{fig:velo}, middle-right and Fig.~\ref{fig:PVmod}, bottom).

The errors on the rotation curve have been estimated as
$\sigma^{2}_{\rm{rot}} = \sigma^{2}_{\rm{fit}} + \sigma^{2}_{\rm{asym}}$, where
$\sigma^{2}_{\rm{fit}}$ is the formal error given by the tilted-ring fit and $\sigma^{2}_{\rm{asym}}$
is an additional uncertainty due to the asymmetries between the approaching and receding sides,
that is estimated as $\sigma{_{\rm{asym}}} = (V_{\rm{rot, app}} - V_{\rm{rot, rec}})/4$
\citep{Swaters2009}. We point out that the observed \hi line profiles are quite broad and,
therefore, it is very difficult to trace the rotation curve in the innermost parts. We
compared models that have an inner solid-body rise and flatten at different radii and
estimated that the rotation curve must flatten between $\sim$0.3 and $\sim$0.6 kpc, giving
an inner rotation-velocity gradient between $\sim$35 and $\sim$65 km~s$^{-1}$~kpc$^{-1}$.
We adopted an intermediate value for the rotation-velocity gradient.

\begin{figure}[tbp]
\centering
\includegraphics[width=6.5cm, angle=-90]{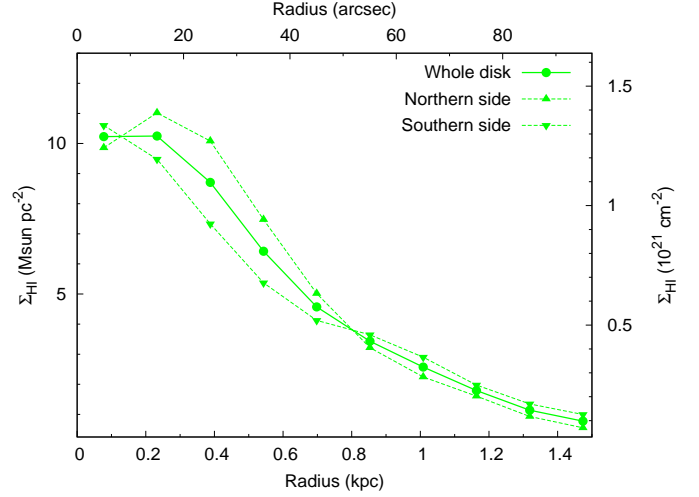}
\caption{\hi surface density profiles (inclination corrected), derived separately for the
northern (up-triangles) and southern (down-triangles) sides and for the entire galaxy (dots).}
\label{fig:HIdens}
\end{figure}

In UGC~4483 the \hi velocity dispersion $\sigma_{\hi}$ is only a factor $\sim$2-3 smaller than
the observed rotation velocity $V_{\rm{rot}}$. Thus, in order to trace the gravitational potential,
the rotation curve has to be corrected for pressure support. We calculated the asymmetric-drift
correction following \citet{Meurer1996}. We assumed that the \hi disk has constant scale-height
and velocity dispersion, and fitted the \hi surface density profile (Fig.~\ref{fig:HIdens})
with the Gaussian function $\Sigma_{\hi}\left(R\right) =\Sigma_{0} \times \exp {\left(-R^2 / 2 s^2 \right)}$,
obtaining $\Sigma_{0} = 10.5$ $M_{\odot}$~pc$^{-2}$ and $s = 580$ pc. The circular velocity $V_{\rm{circ}}$,
corrected for asymmetric-drift, is thus given by $V_{\rm{circ}}^2 = V_{\rm{rot}}^2 + \sigma_{\hi}^2 (R^2/s^2)$.
Using the 3D models, we can constrain the mean velocity dispersion between $\sim$6 and $\sim$10 km~s$^{-1}$.
We assumed the intermediate value of 8~km~s$^{-1}$. The asymmetric-drift correction is significant
only in the outer parts (see Fig. \ref{fig:asymDrift}). Depending on the assumed value of $\sigma_{\hi}$,
the correction at the last measured points may be between $\sim$2 and $\sim$8 km~s$^{-1}$, giving
a dynamical mass between about 1 and 2 $\times10^{8}$ $M_{\odot}$.

In the above analysis, we used the centre and the P.A. derived from the stellar and \hi
morphologies. However, the differences between the observed and the best-model velocity fields 
(Fig. \ref{fig:residuals}, top-right) show a systematic pattern that may point to a different location of the
dynamical centre (cf.  \citealt{Warner1973}, their Fig. 8). Thus, we built further models by changing the
position of the centre and found that a shift of $\sim$14$''$ ($\sim$220~pc) towards the North-West
reduces the pattern significantly. Moreover, a disk-model with an off-set centre and a P.A.$\simeq$10$^{\circ}$
can reproduce the observed velocity field without radial motions (Fig. \ref{fig:residuals}, bottom). 
However, these values for the centre and the P.A. are in marked contrast with the observed stellar and 
\hi morphologies. In particular, the distribution
of the old stars (which constitute an important fraction of the dynamical mass in the inner parts,
see Sect.~\ref{sec:dyn}) is remarkably symmetric and consistent with the centre and P.A. previously 
adopted (see Fig.~6 of \citealt{Dolphin2001} and Fig.~9 of \citealt{Izotov2002}). 
Thus, in the following, we keep the dynamical centre coincident with the optical one.
This has no significant effects on the derived rotation curve.

\begin{figure}[tbp]
\centering
\includegraphics[width=6.5cm, angle=-90]{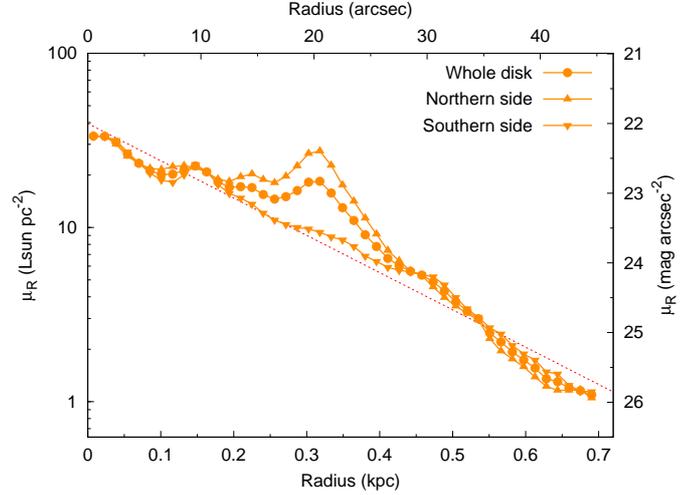}
\caption{R-band surface brightness profiles, derived separately for the northern
(up-triangles) and southern (down-triangles) sides and for the entire galaxy (dots).
The profiles are not corrected for internal extinction and inclination.
The red-dotted line shows an exponential fit to the southern profile,
giving $\mu_{0} = 22.0$ mag~arcsec$^{-2}$
($\sim40$ $L{\odot}$~pc$^{-2}$) and $R_{0} = 13.1'' \pm 0.1''$ ($\sim 203$~pc).}
\label{fig:SBprof}
\end{figure}

\subsection{Mass models \label{sec:dyn}}

\begin{figure*}[tbp]
\centering
\includegraphics[width=7.5cm, angle=-90]{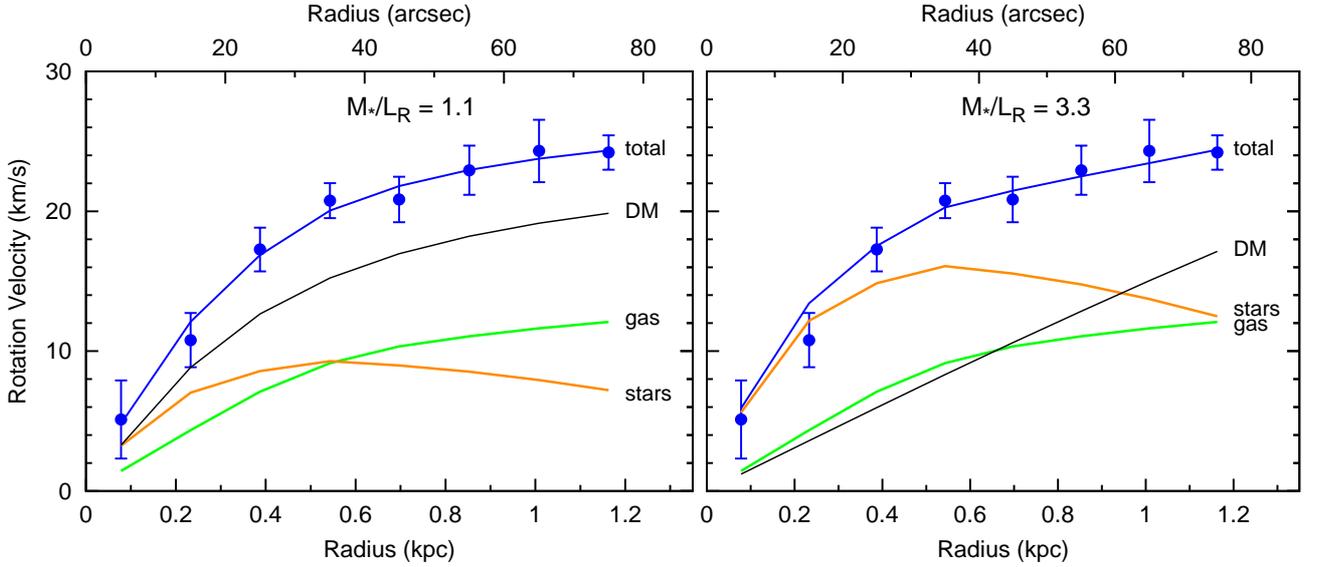}
\caption{Rotation curve decompositions with the $M_{*}/L_{\rm{R}}$ (= 1.1) obtained
from the study of the resolved stellar populations (\textit{left}) and with the
$M_{*}/L_{\rm{R}}$ (= 3.3) from the maximum-disk solution (\textit{right}). Dots show
the observed rotation curve. Lines show the gravitational
contribution of the gas, stars, and dark matter and the resulting rotation curve.}
\label{fig:mass}
\end{figure*}
In Sect. \ref{sec:kin}, we showed that UGC~4483 has a steeply-rising and flat
rotation curve, indicating that there is a strong central concentration of mass.
To determine the relative contributions of luminous and dark matter to the
gravitational potential, we built mass models following \citet{Begeman1987}.

The gravitational contribution of the stars was calculated using a new approach: we
estimated the stellar mass-to-light ratio $M_{*}/L$ using the HST results on the resolved
stellar populations. Surface brightness profiles were derived from a sky-subtracted R-band
image (Fig.~\ref{fig:mosaic}, top-left), azimuthally averaging over a set of ellipses
defined by $(x_{0}, y_{0})$, $i$, and P.A (see table~\ref{tab:prop}). The sky-background
was determined by masking the sources in the field and fitting a 2D polynomial to the masked
image. The surface brightness profiles were not corrected for internal extinction, as UGC~4483
is extremely metal-poor \citep{Skillman1994} and the dust content is likely to be low, but
were corrected for Galactic extinction assuming $A_{\rm{R}}=0.09$ \citep{Schlegel1998}.
Figure \ref{fig:SBprof} shows the surface brightness profiles derived for the entire
galaxy and for the southern (approaching) and northern (receding) halves separately.
At $R\simeq20''$, the northern side of the galaxy is $\sim$1 mag brighter than the southern
one, owing to the recent starburst. \citet{Dolphin2001} and \citet{Izotov2002} resolved
UGC~4483 into single stars and showed that the young stellar populations are mostly concentrated
to the North. The mass in young stars, however, constitutes only $\sim$10$\%$ of the total stellar
mass: the SFH derived by \citet{McQuinn2010} (modelling the color-magnitude diagram) implies that
the stellar mass formed in the last 500 Myr is $\sim$0.15$\times$10$^{7}$~$M_{\odot}$, whereas
the stellar mass formed more than 500 Myr ago is $\sim$1.31$\times$10$^{7}$~$M_{\odot}$.
Thus, to calculate the gravitational contribution of the stars, we used the southern
surface brightness profile, as it provides a better approximation of the stellar mass surface
density. Assuming that $30\%\pm10\%$ of the gas is returned to the inter stellar medium by
supernovae and stellar winds, the total stellar mass is $(1.0 \pm 0.3) \times 10^{7}$~$M_{\odot}$
(cf. \citealt{McQuinn2010b}), giving a stellar mass-to-light ratio $M_{*}/L_{\rm{R}}=1.1\pm0.3$.
The southern surface brightness profile can be fitted by an exponential-law with
$\mu_{0} = 22.0 \pm 0.1$ mag~arcsec$^{-2}$ ($\sim40$ $L_{\odot}$~pc$^{-2}$) and $R_{0} = 13.1'' \pm 0.1''$
($\sim 203$~pc). We assumed that the stars are located in a disk with vertical density distribution given
by $\rho(\rm{z})=\rm{sech}^{2}(z/z_{0})$ \citep{vanDerKruit1981} with z$_{0} \simeq 0.5 R_{0} \simeq 100$~pc.

The gravitational contribution of the gaseous disk was calculated using the azimuthally-averaged
\hi surface density profile, which was derived from the total \hi map. Figure \ref{fig:HIdens}
shows the \hi density profiles derived by azimuthally-averaging over the entire galaxy and over
the southern (approaching) and northern (receding) sides separately. There is a difference of the
order of $\sim$2-3 $M_{\odot}$~pc$^{-2}$ between the two halves, that has a small effect on the
resulting gravitational contribution. In agreement with the models in Sect.~\ref{sec:kin}, we
assumed an exponential vertical distribution with z$_{0}=100$~pc. The mass of the \hi disk was
calculated from the total \hi flux (see table \ref{tab:prop}) and multiplied by a factor
of 1.32 to take into account the presence of helium. Molecular gas was not explicitly
considered in the mass model because its amount is unknown.

For the dark matter distribution, we assumed a pseudo-isothermal halo described
by the equation
\begin{equation} \label{eq:ISO}
 \rho_{\rm{ISO}}(r) = \frac{\rho_{\rm{0}}}{1 + (r/r_{\rm{c}})^{2}} ,
\end{equation}
where the central density $\rho_{\rm{0}}$ and the core radius $r_{\rm{c}}$
are both free parameters of the mass model.

Figure \ref{fig:mass} (left) shows the rotation curve decomposition assuming that $M_{*}/L_{\rm{R}}$=1.1,
as is found by integrating the galaxy SFH \citep{McQuinn2010, McQuinn2010b}. The gravitational contribution
of the stars $V_{*}$ is $\sim$50$\%$ of the circular velocity $V_{\rm{circ}}$ at $R_{\rm{peak}} =
2.2 R_{0} \simeq 450$ pc. The parameters of the halo are: $\rho_{0}= (101 \pm 20) \times 10^{-3}$~$M_{\odot}$~pc$^{-3}$
and $r_{\rm{c}}=0.34 \pm 0.05$~kpc. These values are comparable with those found by \citet{Swaters2011} for
a sample of 18 dIrrs (assuming that $M_{*}/L_{\rm{R}}$=1); the halo of UGC~4483, however, is one with
the smallest core radius and highest central density.
As discussed at the end of Sect~\ref{sec:HIkin}, we applied the asymmetric-drift correction assuming
a constant velocity dispersion $\sigma_{\hi}=8$~km~s$^{-1}$. Different values of $\sigma_{\hi}$ would
change the circular velocity at large radii and thus give slightly different values for the dark matter
halo. In particular, considering that $\sigma_{\hi}$ may be between 6 and 8 km s$^{-1}$, the
dynamical mass within the last measured point of the rotation curve may be between 1 and 2 $\times 10^{8}$~$M_{\odot}$
and the baryon fraction between $\sim$43$\%$ and $\sim$22$\%$, respectively. In any case,
baryons (gas and stars) constitute a relevant fraction of the total mass.

A stellar disk is defined to be maximum if $F_{*} = V_{*}/V_{\rm{circ}} = 0.85 \pm 0.10$ at
$R_{\rm{peak}}$ (\citealt{Sackett1997, Bershady2011}). In our case $F_{*} = 0.5 \pm 0.1$; the error takes
into account the uncertainties on the stellar mass, the gas-recycling efficiency and the rotation velocity.
This is in line with the results of the DiskMass survey \citep{Bershady2011}, that have measured the
stellar velocity dispersion in a sample of spiral galaxies and found that stellar disks typically have
$F_{*} \sim 0.5$. We point out that the colour-magnitude diagrams of the resolved stellar populations
provide the most direct method to quantify stellar masses, as the results depend only slightly on the assumed
evolutionary tracks, metallicity, and sampling of the SFH (e.g. \citealt{Annibali2003}). The initial mass
function (IMF), instead, may have a stronger effect: \citet{McQuinn2010} assumed a single-slope Salpeter
from 0.1 to 100 $M_{\odot}$ and a binary fraction of 35$\%$. If the Chabrier and/or Kroupa IMFs are assumed,
the stellar mass would systematically decrease by, respectively, $\sim$25$\%$ and $\sim$30$\%$
and the gas-recycling efficiency may go up to $\sim$50$\%$, further decreasing the stellar mass.

We note that the molecular gas component has been neglected so far. UGC~4483 is undetected in
the CO line with an upper limit of 0.195 K~km~s$^{-1}$ within the inner 55$''$ ($\sim$800~pc) 
\citep{Taylor1998}. We extrapolated the relation between the CO-to-H$_{2}$ conversion factor
and metallicity from \citet{Boselli2002} down to the metallicity of UGC~4483 and obtained a
corresponding upper limit on the H$_{2}$ mass of $M_{\rm{H}_{2}}\lesssim10^7$ $M_{\odot}$.
Alternatively, we estimated the H$_{2}$ mass from the empirical relation between SFR and molecular
mass, as determined by \citet{Leroy2008}, although this relation may not apply to starburst galaxies.
Assuming that SFR = 0.01 $M_{\odot}$~yr$^{-1}$ \citep{McQuinn2010}, we derived
$M_{\rm{H}_{2}}\simeq1.9 \times 10^{7}$~$M_{\odot}$. Both estimates indicate that the molecular
gas may be dynamically important and comparable in mass to the stellar component, implying a full
maximum-disk situation, as we now discuss.

Figure \ref{fig:mass} (right) shows the maximum-disk solution \citep{Sancisi1987}. The stellar contribution
can explain the inner parts of the rotation curve if $M_{*}/L_{\rm{R}}\simeq3.3$, i.e. if the stellar mass is
$\sim$3 times higher than obtained from the study of the resolved stellar populations. This mass may be
provided by molecules, if they are distributed in a way similar to the stars. We also find a good fit by fixing
$M_{*}/L_{\rm{R}}=1.1$ and scaling the \hi contribution by a factor of $\sim$5, similarly to the results
of \citealt{Hoekstra2001} for other gas-rich galaxies. These results suggest that the distribution of the
dynamical mass is closely coupled to that of the baryonic mass (see e.g. \citealt{Sancisi2004} and
\citealt{Swaters2011}). We also checked the position of UGC~4483 on the baryonic Tully-Fisher relation and
found that it follows the correlation within the observed scatter (cf. \citealt{McGaugh2011}).

\begin{figure}[tbp]
\centering
\includegraphics[width=13cm, angle=-90]{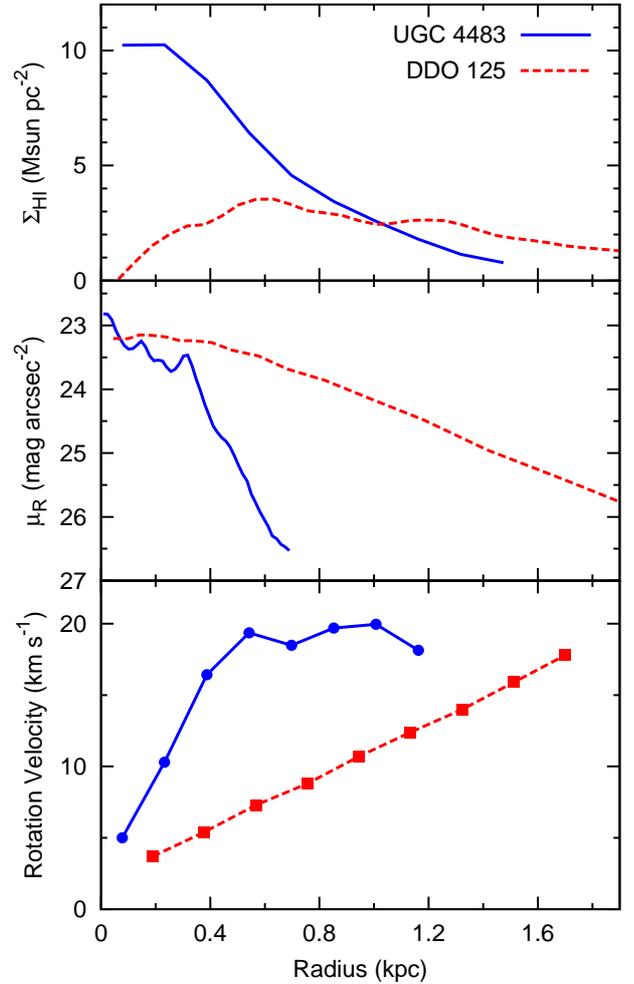}
\caption{Comparison between UGC~4483 (blue line) and the dwarf irregular galaxy
DDO~125 (UGC~7577) (red line). \textit{Top:} HI surface density profiles.
\textit{Middle:} R-band surface brightness profiles, corrected for the inclination
assuming that the disk is transparent and using the ellipticity from \citet{Swaters2002b}.
\textit{Bottom:} HI rotation curves (not corrected for asymmetric drift).}
\label{fig:compa}
\end{figure}
Finally, we considered the predictions of the MOdified Newtonian Dynamics (MOND) (\citealt{Milgrom1983},
see \citealt{Famaey2011} for a review). Using UGC~4483, we can test MOND without any free parameter,
as the value of $M_{*}/L$ is provided by the colour-magnitude diagram of the resolved stellar populations
and the distance is well determined from the tip of the red giant branch \citep{Dolphin2001, Izotov2002}.
We found that MOND systematically over-predicts the observed rotation curve by $\sim$5-6 km~s$^{-1}$.
The discrepancy does not strongly depend on the assumed $M_{*}/L$, as it is mainly driven by the atomic gas
content, and it would further increase if molecules are also considered in the mass model. However, the
discrepancy disappears if the inclination of the \hi disk is assumed to be $\sim$43$^{\circ}$ instead of
58$^{\circ}$ (as is derived from the optical image). This possibility cannot be ruled out by the present
observations, because it is not possible to determine $i$ neither from the total \hi map (as the \hi
distribution is strongly lopsided) nor from the velocity field.

\section{Discussion \label{sec:discuss}}

\begin{figure*}[tbp]
\centering
\includegraphics[width=8.5cm, angle=-90]{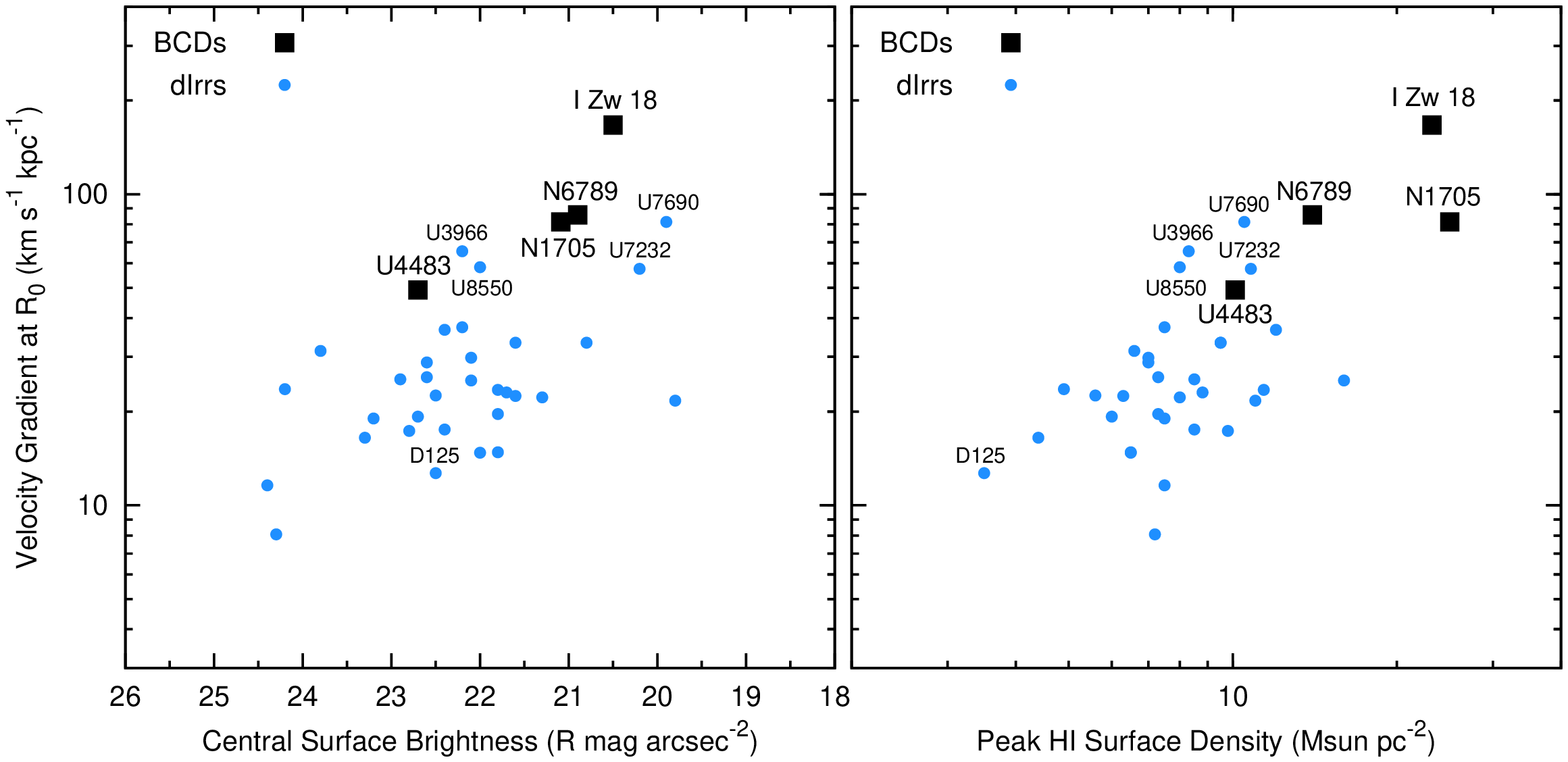}
\caption{Comparison between the properties of BCDs (squares) and of dIrrs (dots).
The dIrrs are taken from \citet{Swaters2009}, selecting the objects with high-quality
rotation curves and $V_{\rm{rot}} < 70$ km~s$^{-1}$ at the last measured point.
The BCDs are UGC~4483 (this work), I~Zw~18 \citep{Lelli2012}, NGC~6789 (Lelli et al. in preparation),
and NGC~1705 \citep{Meurer1998}. Some galaxies are labelled according to their UGC number.
\textit{Left:} R-band central surface brightness (inclination corrected) versus the inner
rotation-velocity gradient $V_{\rm{rot}}(R_{0})/R_{0}$, where $R_{0}$ is the optical scale-length.
\textit{Right:} peak \hi surface density $\Sigma_{\hiV, \, \rm{peak}}$ versus $V_{\rm{rot}}(R_{0})/R_{0}$.
$\Sigma_{\hiV, \, \rm{peak}}$ is derived from the azimuthally-averaged \hi surface density
profiles. See Sect.~\ref{sec:discuss} for details.}
\label{fig:GradPlot}
\end{figure*}
In Sect. \ref{sec:kin}, we showed that UGC~4483 has a steeply-rising rotation curve that flattens
in the outer parts at a velocity of $\sim$18-20 km~s$^{-1}$. As far as we know, this is the
lowest-mass galaxy with a differentially rotating \hi disk. The steep rise of the rotation
curve points to a strong central concentration of mass, which seems to be a characterizing property
of BCDs. Indeed, similar results have been found also for other BCDs, e.g. I~Zw~18 \citep{Lelli2012},
NGC~2537 \citep{Matthews2008}, and NGC~1705 \citep{Meurer1998}. The steeply-rising and flat rotation
curves of these BCDs are remarkable, as dIrrs typically have slowly-rising rotation curves
(e.g. \citealt{Swaters2009}). The terms ``steeply-rising'' and ``slowly-rising'' are referred
to rotation-velocity gradients measured in physical units (km~s$^{-1}$~kpc$^{-1}$) that can be directly
related to the dynamical mass surface densities in $M_{\odot}$~pc$^{-2}$. If these gradients are
expressed in terms of disk scale-lengths, rotation curves of dIrrs rise as steep as those of
spiral galaxies (e.g. \citealt{Swaters2009}).

In \citet{Lelli2012} (Fig. 10), we have illustrated the dynamical difference between
the BCD-prototype I~Zw~18 and a typical dIrr of the same mass.
Similarly, Fig.~\ref{fig:compa} compares UGC~4483 with the dwarf irregular DDO~125 (UGC~7577,
from \citealt{Swaters2009}). Assuming that DDO~125 is at a distance of 2.6~Mpc \citep{Jacobs2009, Dalcanton2009},
the \hi observations of \citet{Swaters2009} have a linear resolution of $\sim$180~pc, similar to
our linear resolution for UGC~4483 ($\sim$155~pc), making it possible to compare \hi surface densities
and velocity gradients. The two galaxies have approximately the same rotation velocity at the last
measured point ($\sim$20 km s$^{-1}$) and dynamical mass ($\sim$1-2$\times$10$^{8}$~$M_{\odot}$),
but their structural properties are very different: i) the \hi surface densities of UGC~4483 are
$\sim$3-4 times higher than those of DDO~125 in the inner regions (top);
ii) the stellar component of UGC~4483 is much more compact than that of DDO~125 (middle);
iii) the rotation curve of UGC~4483 has a steeper rise than the one of DDO~125 and flattens
in the outer parts. These structural and dynamical differences between BCDs and typical dIrrs
must be the key to understand the starburst phenomenon in BCDs. They also shed new light on the
question of the progenitors and descendants of BCDs, as we now discuss.

In Section \ref{sec:dyn}, we showed that the central mass concentration of UGC~4483 cannot be explained
by the newly formed stars or by the concentration of \hiV. Old stars, although not sufficient to account
for the inner rise of the rotation curve, constitute an important fraction of the mass in the inner parts.
Similarly, the central mass concentration of I~Zw~18 may be due to old stars and/or dark matter
(unfortunately, the HST observations of I~Zw~18 do not provide a direct estimate of the mass in old stars,
see \citealt{ContrerasRamos2011}). For both galaxies, the gravitational contribution of molecules is very
uncertain and it is unclear whether they are dynamically important or not. These results imply that
either the progenitors of these BCDs are unusually compact, gas-rich dwarfs, or there must be a
mechanism that leads to a concentration of gas, old stars, and/or dark matter, eventually causing
the starburst. This mechanism may be external (interactions/mergers) and/or internal
(torques from massive star-forming ``clumps'', see \citealt{Elmegreen2012}). It is also clear that,
unless a redistribution of mass takes place, the descendants of BCDs must be compact dwarfs.
Compact dIrrs do exist and have the following properties: i) high-surface-brightness exponential
profiles (see e.g. UGC~7690 and UGC~8550 in \citealt{Swaters2002b}) or low-surface-brightness
exponential profiles with an inner light concentration (see e.g. UGC~6628 and UGC~12632 in \citealt{Swaters2002b}),
and ii) steeply-rising and flat rotation curves (see the same galaxies in \citealt{Swaters2009}). The evolutionary
links outlined here are in line with the results of e.g. \citet{Papaderos1996} and \citet{GilDePaz2005},
which are based on surface photometry of large galaxy samples and indicate that the old stellar component
of BCDs generally has higher central surface brightness and smaller scale-length than typical dIrrs and dSphs/dEs.
We present here two plots that compare the structural and dynamical properties of BCDs and dIrrs.

In order to compare the \textit{dynamics} of BCDs and dIrrs, we use the inner rotation-velocity gradient
$V_{\rm{rot}}(R)/R$ as an estimate of central mass concentration. This is calculated at $R = R_{0}$, where
$R_{0}$ is the galaxy scale-length in the R-band; thus we take into account the different sizes of the systems.
In Fig.~\ref{fig:GradPlot}, $V_{\rm{rot}}(R_{0})/R_{0}$ is plotted versus the central disk surface brightness
$\mu_{0}$ (left) and the peak \hi column density $\Sigma_{\hiV,\, \rm{peak}}$ (right) for a sample of both BCDs and dIrrs.
The BCDs are UGC~4483, I~Zw~18, NGC~1705, and NGC~6789. For I~Zw~18 we used the R-band structural parameters from
\citet{Papaderos2002}, the rotation curve from \citet{Lelli2012}, and a distance of 18.2 Mpc \citep{Aloisi2007}.
For NGC~1705 we used the R-band structural parameters from \citet{GilDePaz2005}, the rotation curve from
\citet{Meurer1998} named ``model DD'' (see their Fig.~10), and a distance of 5.1 Mpc \citep{Tosi2001}.
For NGC~6789 we used the R-band structural parameters and the rotation curve from Lelli et al. (in preparation),
and a distance of 3.6~Mpc \citep{Drozdovsky2001}.
The dIrrs were taken from the sample of \citet{Swaters2009}, by selecting the objects that have high-quality
rotation curves ($q\leq2$, see \citealt{Swaters2009}) and $V_{\rm{rot}} < 70$ km~s$^{-1}$ at the last measured
point. For $\Sigma_{\hiV,\, \rm{peak}}$, we used the peak value of the azimuthally-averaged \hi density profile.
The value of $\Sigma_{\hiV,\, \rm{peak}}$ depends on the linear resolution of the \hi observations, but we checked
that this does not strongly bias our analysis as the trend in Fig.~\ref{fig:GradPlot} does not significantly
change if one considers only galaxies with linear resolution $\lesssim$ 500~pc (corresponding to distances
$\lesssim$ 7 Mpc). We also point out that BCDs usually have clumpy \hi distributions and locally the \hi
column densities can reach even higher values, up to 40-50 $M_{\odot}$~pc$^{-2}$, much higher than in dIrrs.

The two diagrams in figure \ref{fig:GradPlot} show a clear trend: galaxies with a high rotation-velocity
gradient have also high central surface brightness (left) and high peak \hi surface density (right). The BCDs
are in the upper parts of these distributions: they have $V_{\rm{rot}}(R_{0})/R_{0}>40$ km s$^{-1}$ kpc$^{-1}$
whereas dIrrs typically have $V_{\rm{rot}}(R_{0})/R_{0}<40$ km s$^{-1}$ kpc$^{-1}$. However, some dIrrs
(UGC~3966, UGC~8550, UGC~7232, and UGC~7690) have inner rotation-velocity gradients comparable to those
of BCDs and thus, in this respect, are dynamically similar to BCDs. These objects have peak \hi column
densities much smaller than those found in I~Zw~18 and NGC~1705 (Fig. \ref{fig:GradPlot}, right)
and are candidate progenitors/descendants of BCDs. It would be interesting to study the SFHs of
these compact dIrrs and investigate if they experienced a starburst in the recent past.

Finally, we have seen that the asymmetries in the velocity field of UGC~4483
can be described as a global radial motion of 5 km~s$^{-1}$. Radial motions have been found in two other BCDs: NGC~2915
\citep{Elson2011, Elson2011b} and I~Zw~18 \citep{Lelli2012}. \citet{Elson2011} assumed that
the spiral arms in the \hi disk of NGC~2915 are trailing and concluded that the radial motions
are an outflow. For I~Zw~18 and UGC~4483, it is not possible to discriminate between inflow
and outflow, as it is not known which side of the disk is the near one. For both galaxies,
we calculated the timescales associated with the radial motions ($\simeq$~$R_{\hi}/V_{\rm{rad}}$) 
and found that they are of the same order of magnitude of the orbital times ($\simeq$~$2\pi R_{\hi}/V_{\rm{rot}}$).
This suggests that any outflow or inflow must be very recent and possibly associated with
the most recent burst of star-formation ($\sim$10-20~Myr, see \citealt{McQuinn2010} for UGC~4483
and \citealt{Aloisi1999} for I~Zw~18). If the radial motions were an outflow, its kinetic energy
would correspond to only $\sim$1$\%$ of the energy released by supernovae.

\section{Conclusions}

We analysed archival \hi observations of the blue compact dwarf galaxy
UGC~4483 and built model datacubes to investigate its gas kinematics.
Our main results can be summarized as follows:
\begin{itemize}
 \item UGC~4483 has a steeply-rising rotation curve that flattens in the outer parts
       at a velocity of $\sim$20 km s$^{-1}$. This is, to our knowledge, the lowest-mass
       galaxy with a differentially rotating \hi disk. Radial motions
       of $\sim$5 km~s$^{-1}$ may also be present.
 \item The steep rise of the rotation curve indicates that there is a strong central
       concentration of mass. Mass models with a dark matter halo show that \textit{old} stars
       contribute $\sim$50$\%$ of the observed rotation speed at 2.2 disk scale-lengths.
       Baryons (gas and stars) constitute an important fraction of the total dynamical mass.
       These conclusions are based on the stellar
       mass obtained from the color-magnitude diagram of the resolved stellar populations.
 \item The maximum-disk solution requires a stellar mass 3 times higher than observed,
       that could be provided by molecules. A good solution is also found by scaling the
       \hi contribution by a factor of $\sim$5. These results suggest that the distribution
       of the dynamical mass is closely coupled to that of the baryons.
\end{itemize}
UGC~4483, together with other BCDs like I~Zw~18 and NGC~1705, appears structurally different
from typical dIrrs in terms of \hi distribution, stellar distribution, and dynamics.
In particular, a central concentration of mass (gas, stars, and dark matter) seems to be a
characterizing property of BCDs. This implies that the starburst is closely related
with the gravitational potential and the \hi concentration. Our results also
suggest that the progenitors/descendants of BCDs must be compact dwarf galaxies, unless
a redistribution of mass (both luminous and dark) takes place before/after the starbursting phase.

\begin{acknowledgements}
We thank Rob Swaters for stimulating comments.
We are grateful to F. Annibali and M. Tosi for helpful discussions about the stellar
populations of BCDs. We thank the ISSI (Bern) for support of the team ``Defining
the full life-cycle of dwarf galaxy evolution: the Local Universe as a template''.
\end{acknowledgements}

\bibliographystyle{aa}
\bibliography{U4483bib.bib}

\begin{thebibliography}{59}
\expandafter\ifx\csname natexlab\endcsname\relax\def\natexlab#1{#1}\fi

\bibitem[{{Aloisi} {et~al.}(2007){Aloisi}, {Clementini}, {Tosi}, {Annibali},
  {Contreras}, {Fiorentino}, {Mack}, {Marconi}, {Musella}, {Saha}, {Sirianni},
  \& {van der Marel}}]{Aloisi2007}
{Aloisi}, A., {Clementini}, G., {Tosi}, M., {et~al.} 2007, \apjl, 667, L151

\bibitem[{{Aloisi} {et~al.}(1999){Aloisi}, {Tosi}, \& {Greggio}}]{Aloisi1999}
{Aloisi}, A., {Tosi}, M., \& {Greggio}, L. 1999, \aj, 118, 302

\bibitem[{{Annibali} {et~al.}(2003){Annibali}, {Greggio}, {Tosi}, {Aloisi}, \&
  {Leitherer}}]{Annibali2003}
{Annibali}, F., {Greggio}, L., {Tosi}, M., {Aloisi}, A., \& {Leitherer}, C.
  2003, \aj, 126, 2752

\bibitem[{{Begeman}(1987)}]{Begeman1987}
{Begeman}, K.~G. 1987, PhD thesis, , Kapteyn Institute, (1987)

\bibitem[{{Bershady} {et~al.}(2011){Bershady}, {Martinsson}, {Verheijen},
  {Westfall}, {Andersen}, \& {Swaters}}]{Bershady2011}
{Bershady}, M.~A., {Martinsson}, T.~P.~K., {Verheijen}, M.~A.~W., {et~al.}
  2011, \apjl, 739, L47

\bibitem[{{Binney} \& {Merrifield}(1998)}]{BinneyMerrifield1998}
{Binney}, J. \& {Merrifield}, M. 1998, {Galactic astronomy}, ed. {Binney, J.~\&
  Merrifield, M.}

\bibitem[{{Boselli} {et~al.}(2002){Boselli}, {Lequeux}, \&
  {Gavazzi}}]{Boselli2002}
{Boselli}, A., {Lequeux}, J., \& {Gavazzi}, G. 2002, \aap, 384, 33

\bibitem[{{Bosma}(1978)}]{Bosma1978}
{Bosma}, A. 1978, PhD thesis, PhD Thesis, Groningen Univ., (1978)

\bibitem[{{Briggs}(1995)}]{Briggs1995}
{Briggs}, D.~S. 1995, in Bulletin of the American Astronomical Society,
  Vol.~27, Bulletin of the American Astronomical Society, 1444--+

\bibitem[{{Contreras Ramos} {et~al.}(2011){Contreras Ramos}, {Annibali},
  {Fiorentino}, {Tosi}, {Aloisi}, {Clementini}, {Marconi}, {Musella}, {Saha},
  \& {van der Marel}}]{ContrerasRamos2011}
{Contreras Ramos}, R., {Annibali}, F., {Fiorentino}, G., {et~al.} 2011, \apj,
  739, 74

\bibitem[{{Dalcanton} {et~al.}(2009){Dalcanton}, {Williams}, {Seth}, {Dolphin},
  {Holtzman}, {Rosema}, {Skillman}, {Cole}, {Girardi}, {Gogarten},
  {Karachentsev}, {Olsen}, {Weisz}, {Christensen}, {Freeman}, {Gilbert},
  {Gallart}, {Harris}, {Hodge}, {de Jong}, {Karachentseva}, {Mateo}, {Stetson},
  {Tavarez}, {Zaritsky}, {Governato}, \& {Quinn}}]{Dalcanton2009}
{Dalcanton}, J.~J., {Williams}, B.~F., {Seth}, A.~C., {et~al.} 2009, \apjs,
  183, 67

\bibitem[{{Dolphin} {et~al.}(2001){Dolphin}, {Makarova}, {Karachentsev},
  {Karachentseva}, {Geisler}, {Grebel}, {Guhathakurta}, {Hodge}, {Sarajedini},
  \& {Seitzer}}]{Dolphin2001}
{Dolphin}, A.~E., {Makarova}, L., {Karachentsev}, I.~D., {et~al.} 2001, \mnras,
  324, 249

\bibitem[{{Drozdovsky} {et~al.}(2001){Drozdovsky}, {Schulte-Ladbeck}, {Hopp},
  {Crone}, \& {Greggio}}]{Drozdovsky2001}
{Drozdovsky}, I.~O., {Schulte-Ladbeck}, R.~E., {Hopp}, U., {Crone}, M.~M., \&
  {Greggio}, L. 2001, \apjl, 551, L135

\bibitem[{{Elmegreen} {et~al.}(2012){Elmegreen}, {Zhang}, \&
  {Hunter}}]{Elmegreen2012}
{Elmegreen}, B.~G., {Zhang}, H., \& {Hunter}, D. 2012, ArXiv e-prints

\bibitem[{{Elson} {et~al.}(2011{\natexlab{a}}){Elson}, {de Blok}, \&
  {Kraan-Korteweg}}]{Elson2011}
{Elson}, E.~C., {de Blok}, W.~J.~G., \& {Kraan-Korteweg}, R.~C.
  2011{\natexlab{a}}, \mnras, 411, 200

\bibitem[{{Elson} {et~al.}(2011{\natexlab{b}}){Elson}, {de Blok}, \&
  {Kraan-Korteweg}}]{Elson2011b}
{Elson}, E.~C., {de Blok}, W.~J.~G., \& {Kraan-Korteweg}, R.~C.
  2011{\natexlab{b}}, \mnras, 415, 323

\bibitem[{{Famaey} \& {McGaugh}(2011)}]{Famaey2011}
{Famaey}, B. \& {McGaugh}, S. 2011, ArXiv e-prints

\bibitem[{{Fraternali} {et~al.}(2002){Fraternali}, {van Moorsel}, {Sancisi}, \&
  {Oosterloo}}]{Fraternali2002}
{Fraternali}, F., {van Moorsel}, G., {Sancisi}, R., \& {Oosterloo}, T. 2002,
  \aj, 123, 3124

\bibitem[{{Gallagher} \& {Hunter}(1987)}]{Gallagher1987}
{Gallagher}, III, J.~S. \& {Hunter}, D.~A. 1987, \aj, 94, 43

\bibitem[{{Gil de Paz} \& {Madore}(2005)}]{GilDePaz2005}
{Gil de Paz}, A. \& {Madore}, B.~F. 2005, \apjs, 156, 345

\bibitem[{{Gil de Paz} {et~al.}(2003){Gil de Paz}, {Madore}, \&
  {Pevunova}}]{GilDePaz2003}
{Gil de Paz}, A., {Madore}, B.~F., \& {Pevunova}, O. 2003, \apjs, 147, 29

\bibitem[{{Hoekstra} {et~al.}(2001){Hoekstra}, {van Albada}, \&
  {Sancisi}}]{Hoekstra2001}
{Hoekstra}, H., {van Albada}, T.~S., \& {Sancisi}, R. 2001, \mnras, 323, 453

\bibitem[{{H{\"o}gbom}(1974)}]{Hogbom1974}
{H{\"o}gbom}, J.~A. 1974, \aaps, 15, 417

\bibitem[{{Izotov} \& {Thuan}(2002)}]{Izotov2002}
{Izotov}, Y.~I. \& {Thuan}, T.~X. 2002, \apj, 567, 875

\bibitem[{{Jacobs} {et~al.}(2009){Jacobs}, {Rizzi}, {Tully}, {Shaya},
  {Makarov}, \& {Makarova}}]{Jacobs2009}
{Jacobs}, B.~A., {Rizzi}, L., {Tully}, R.~B., {et~al.} 2009, \aj, 138, 332

\bibitem[{{Lelli} {et~al.}(2010){Lelli}, {Fraternali}, \&
  {Sancisi}}]{Lelli2010}
{Lelli}, F., {Fraternali}, F., \& {Sancisi}, R. 2010, \aap, 516, A11+

\bibitem[{{Lelli} {et~al.}(2012){Lelli}, {Verheijen}, {Fraternali}, \&
  {Sancisi}}]{Lelli2012}
{Lelli}, F., {Verheijen}, M., {Fraternali}, F., \& {Sancisi}, R. 2012, \aap,
  537, A72

\bibitem[{{Leroy} {et~al.}(2008){Leroy}, {Walter}, {Brinks}, {Bigiel}, {de
  Blok}, {Madore}, \& {Thornley}}]{Leroy2008}
{Leroy}, A.~K., {Walter}, F., {Brinks}, E., {et~al.} 2008, \aj, 136, 2782

\bibitem[{{Lo} {et~al.}(1993){Lo}, {Sargent}, \& {Young}}]{Lo1993}
{Lo}, K.~Y., {Sargent}, W.~L.~W., \& {Young}, K. 1993, \aj, 106, 507

\bibitem[{{Martin}(1996)}]{Martin1996}
{Martin}, C.~L. 1996, \apj, 465, 680

\bibitem[{{Matthews} \& {Uson}(2008)}]{Matthews2008}
{Matthews}, L.~D. \& {Uson}, J.~M. 2008, \aj, 135, 291

\bibitem[{{McGaugh}(2011)}]{McGaugh2011}
{McGaugh}, S.~S. 2011, Physical Review Letters, 106, 121303

\bibitem[{{McQuinn} {et~al.}(2010{\natexlab{a}}){McQuinn}, {Skillman},
  {Cannon}, {Dalcanton}, {Dolphin}, {Hidalgo-Rodr{\'{\i}}guez}, {Holtzman},
  {Stark}, {Weisz}, \& {Williams}}]{McQuinn2010}
{McQuinn}, K.~B.~W., {Skillman}, E.~D., {Cannon}, J.~M., {et~al.}
  2010{\natexlab{a}}, \apj, 721, 297

\bibitem[{{McQuinn} {et~al.}(2010{\natexlab{b}}){McQuinn}, {Skillman},
  {Cannon}, {Dalcanton}, {Dolphin}, {Hidalgo-Rodr{\'{\i}}guez}, {Holtzman},
  {Stark}, {Weisz}, \& {Williams}}]{McQuinn2010b}
{McQuinn}, K.~B.~W., {Skillman}, E.~D., {Cannon}, J.~M., {et~al.}
  2010{\natexlab{b}}, \apj, 724, 49

\bibitem[{{Meurer} {et~al.}(1996){Meurer}, {Carignan}, {Beaulieu}, \&
  {Freeman}}]{Meurer1996}
{Meurer}, G.~R., {Carignan}, C., {Beaulieu}, S.~F., \& {Freeman}, K.~C. 1996,
  \aj, 111, 1551

\bibitem[{{Meurer} {et~al.}(1998){Meurer}, {Staveley-Smith}, \&
  {Killeen}}]{Meurer1998}
{Meurer}, G.~R., {Staveley-Smith}, L., \& {Killeen}, N.~E.~B. 1998, \mnras,
  300, 705

\bibitem[{Milgrom(1983)}]{Milgrom1983}
Milgrom, M. 1983, MNRAS, 270, 365

\bibitem[{{Papaderos} {et~al.}(2002){Papaderos}, {Izotov}, {Thuan}, {Noeske},
  {Fricke}, {Guseva}, \& {Green}}]{Papaderos2002}
{Papaderos}, P., {Izotov}, Y.~I., {Thuan}, T.~X., {et~al.} 2002, \aap, 393, 461

\bibitem[{{Papaderos} {et~al.}(1996){Papaderos}, {Loose}, {Fricke}, \&
  {Thuan}}]{Papaderos1996}
{Papaderos}, P., {Loose}, H., {Fricke}, K.~J., \& {Thuan}, T.~X. 1996, \aap,
  314, 59

\bibitem[{{Sackett}(1997)}]{Sackett1997}
{Sackett}, P.~D. 1997, \apj, 483, 103

\bibitem[{{Sancisi}(2004)}]{Sancisi2004}
{Sancisi}, R. 2004, in IAU Symposium, Vol. 220, Dark Matter in Galaxies, ed.
  {S.~Ryder, D.~Pisano, M.~Walker, \& K.~Freeman}, 233--+

\bibitem[{{Sancisi} \& {van Albada}(1987)}]{Sancisi1987}
{Sancisi}, R. \& {van Albada}, T.~S. 1987, in IAU Symposium, Vol. 117, Dark
  matter in the universe, ed. {J.~Kormendy \& G.~R.~Knapp}, 67--80

\bibitem[{{Schlegel} {et~al.}(1998){Schlegel}, {Finkbeiner}, \&
  {Davis}}]{Schlegel1998}
{Schlegel}, D.~J., {Finkbeiner}, D.~P., \& {Davis}, M. 1998, \apj, 500, 525

\bibitem[{{Searle} \& {Sargent}(1972)}]{Searle1972}
{Searle}, L. \& {Sargent}, W.~L.~W. 1972, \apj, 173, 25

\bibitem[{{Skillman} {et~al.}(1994){Skillman}, {Televich}, {Kennicutt},
  {Garnett}, \& {Terlevich}}]{Skillman1994}
{Skillman}, E.~D., {Televich}, R.~J., {Kennicutt}, Jr., R.~C., {Garnett},
  D.~R., \& {Terlevich}, E. 1994, \apj, 431, 172

\bibitem[{{Swaters} \& {Balcells}(2002)}]{Swaters2002b}
{Swaters}, R.~A. \& {Balcells}, M. 2002, \aap, 390, 863

\bibitem[{{Swaters} {et~al.}(2009){Swaters}, {Sancisi}, {van Albada}, \& {van
  der Hulst}}]{Swaters2009}
{Swaters}, R.~A., {Sancisi}, R., {van Albada}, T.~S., \& {van der Hulst}, J.~M.
  2009, \aap, 493, 871

\bibitem[{{Swaters} {et~al.}(2011){Swaters}, {Sancisi}, {van Albada}, \& {van
  der Hulst}}]{Swaters2011}
{Swaters}, R.~A., {Sancisi}, R., {van Albada}, T.~S., \& {van der Hulst}, J.~M.
  2011, \apj, 729, 118

\bibitem[{{Taylor} {et~al.}(1995){Taylor}, {Brinks}, {Grashuis}, \&
  {Skillman}}]{Taylor1995}
{Taylor}, C.~L., {Brinks}, E., {Grashuis}, R.~M., \& {Skillman}, E.~D. 1995,
  \apjs, 99, 427

\bibitem[{{Taylor} {et~al.}(1998){Taylor}, {Kobulnicky}, \&
  {Skillman}}]{Taylor1998}
{Taylor}, C.~L., {Kobulnicky}, H.~A., \& {Skillman}, E.~D. 1998, \aj, 116, 2746

\bibitem[{{Tosi}(2009)}]{Tosi2009}
{Tosi}, M. 2009, in IAU Symposium, Vol. 258, IAU Symposium, ed. {E.~E.~Mamajek,
  D.~R.~Soderblom, \& R.~F.~G.~Wyse}, 61--72

\bibitem[{{Tosi} {et~al.}(2001){Tosi}, {Sabbi}, {Bellazzini}, {Aloisi},
  {Greggio}, {Leitherer}, \& {Montegriffo}}]{Tosi2001}
{Tosi}, M., {Sabbi}, E., {Bellazzini}, M., {et~al.} 2001, \aj, 122, 1271

\bibitem[{van~der Hulst {et~al.}(1992)van~der Hulst, Terlouw, Begeman, Zwitser,
  \& Roelfsema}]{vanderHulst1992}
van~der Hulst, J., Terlouw, J., Begeman, K., Zwitser, W., \& Roelfsema, P.
  1992, in ASP Conf. Ser. 25, ed. D.~M. Worall, C.~Biemesderfer, \& J.~Barnes,
  San Francisco: ASP, 131

\bibitem[{{van der Kruit} \& {Searle}(1981)}]{vanDerKruit1981}
{van der Kruit}, P.~C. \& {Searle}, L. 1981, \aap, 95, 105

\bibitem[{{van Zee} \& {Haynes}(2006)}]{vanZee2006}
{van Zee}, L. \& {Haynes}, M.~P. 2006, \apj, 636, 214

\bibitem[{{van Zee} {et~al.}(2001){van Zee}, {Salzer}, \&
  {Skillman}}]{vanZee2001}
{van Zee}, L., {Salzer}, J.~J., \& {Skillman}, E.~D. 2001, \aj, 122, 121

\bibitem[{{van Zee} {et~al.}(1998){van Zee}, {Skillman}, \&
  {Salzer}}]{vanZee1998b}
{van Zee}, L., {Skillman}, E.~D., \& {Salzer}, J.~J. 1998, \aj, 116, 1186

\bibitem[{Verheijen \& Sancisi(2001)}]{Verheijen2001}
Verheijen, M. \& Sancisi, R. 2001, A\&A, 370, 765

\bibitem[{{Warner} {et~al.}(1973){Warner}, {Wright}, \& {Baldwin}}]{Warner1973}
{Warner}, P.~J., {Wright}, M.~C.~H., \& {Baldwin}, J.~E. 1973, \mnras, 163, 163

\end{thebibliography}
\end{document}